\ifdefined\XeTeXversion\else\ifdefined\pdfoutput\pdfoutput=1\fi\fi

\documentclass[conference]{IEEEtran}
\IEEEoverridecommandlockouts

\usepackage[T1]{fontenc}
\usepackage{graphicx}
\usepackage{amsmath}
\usepackage{booktabs}
\usepackage{microtype}
\usepackage{url}
\usepackage[hidelinks]{hyperref}

\newcommand{\CovProteins}{8{,}510}
\newcommand{\CovInteractions}{34{,}540}
\newcommand{\CovComponents}{1}
\newcommand{\CovBridges}{2{,}848}
\newcommand{\CovArticulation}{32}
\newcommand{\CovThreshold}{0.2}
\newcommand{\CovThresholdProteins}{3{,}864}
\newcommand{\CovThresholdInteractions}{11{,}320}
\newcommand{\CovSinglePublication}{24{,}344}

\newcommand{\CovSinglePercent}{70}

\newcommand{\CovCliques}{19{,}136}
\newcommand{\CovLargestClique}{9}
\newcommand{\CovLargestCliqueMembers}{GNB2L1, N, ORF6, nsp1, nsp10, nsp12, nsp14, nsp3, nsp5}
\newcommand{\CorrStrict}{0.512}
\newcommand{\CorrEvidence}{0.302}
\newcommand{\CorrLiterature}{0.325}

\newcommand{\HumanProteins}{29{,}104}
\newcommand{\HumanInteractions}{1{,}047{,}820}
\newcommand{\HumanModules}{14}
\newcommand{\HumanSteps}{4}
\newcommand{\HumanLeaf}{53}
\newcommand{\HumanDescentSeconds}{1.9}
\newcommand{\OrganismsSurveyed}{41}
\newcommand{\OrganismsStructural}{10}
\newcommand{\HumanPublications}{41{,}218}
\newcommand{\HumanPublicationsDrawn}{16{,}148}
\newcommand{\HumanMethods}{29}
\newcommand{\HumanBiggestPublication}{166{,}915}
\newcommand{\ScreenCut}{500}
\newcommand{\ScreenPublications}{315}
\newcommand{\ScreenMethods}{25}

\newcommand{\NetProteins}{48}
\newcommand{\NetInteractions}{248}
\newcommand{\NetCommunities}{4}

\title{ProLiVis 2.0: Literature-Centric Visualization of Protein--Protein
Interaction Networks, with a Citation-Trust Model for Interaction Evidence}

\author{
\IEEEauthorblockN{Melih Sözdinler\IEEEauthorrefmark{1}\IEEEauthorrefmark{2}, Yalç\i{}n Doksanbir\IEEEauthorrefmark{1}, Gökhan Akp\i{}nar\IEEEauthorrefmark{1}, Ege Aktan\IEEEauthorrefmark{1}}
\IEEEauthorblockA{\IEEEauthorrefmark{1}Emlakjet, R\&D Center, R\&D, 34764 \.{I}stanbul, Turkey \\
Email: melih.sozdinler@emlakjet.com (M.S.); yalcin.doksanbir@emlakjet.com (Y.D.); \\
gokhan.akpinar@emlakjet.com (G.A.); ege.aktan@emlakjet.com (E.A.)}
\IEEEauthorblockA{\IEEEauthorrefmark{2}Computer Engineering, Faculty of Engineering and Natural Sciences, \\
I\c{s}\i{}k University, 34980 \.{I}stanbul, Turkey}
\thanks{\raggedright Corresponding author: Melih Sözdinler (melih.sozdinler@emlakjet.com)\par}
}

\begin{document}
\maketitle

\begin{abstract}
Protein--protein interaction databases record evidence without weighing it. In BioGRID,
an interaction asserted once by a single high-throughput screen and one confirmed by
twenty laboratories across a dozen assays are the same kind of row in the same file.
Tools built on such databases inherit that flattening: they draw every reported
interaction as an edge, and the resulting picture states that two proteins interact
without stating how much anyone should believe it.

We present ProLiVis 2.0, a rewrite of the literature-centric visualization system of
\cite{prolivis1}. It contributes three things. First, a \emph{citation-trust model}
that scores each interaction from seven terms, including a term for the number of
\emph{independent laboratories} behind the supporting publications, obtained by
clustering those publications over shared institutional affiliations; a plain count of
publications cannot distinguish five confirmations from one group publishing five times.
Second, a deterministic reformulation of the center layout, closed-form and
$O(n \log n)$, which replaces the force-directed placement of the original and makes
published figures regenerable from a session manifest. Third, an implementation that
runs entirely in a web browser, with an embedded analytical database, requiring no
installation and uploading no data.

On BioGRID release 5.0.260 restricted to SARS-CoV-2, \CovSinglePublication{} of
\CovInteractions{} reported interactions --- \CovSinglePercent\% --- rest on a single
publication, and raising the trust threshold to $\CovThreshold$ leaves
\CovThresholdInteractions{} of them. That the large majority of a curated interaction
network is unreplicated is a fact no existing view of the database makes visible.
\end{abstract}


\section{Introduction}

BioGRID \cite{biogrid} curates over two million reported protein and genetic
interactions. Each record states that one publication, using one experimental system,
reported that two genes interact. It does not state whether the assay was appropriate,
whether anyone reproduced the result, or whether the reporting publication was ever
cited. Those facts exist in the records --- distributed across many rows --- but no
field holds them, and no standard view surfaces them.

The consequence is familiar to anyone who has drawn such a network. Every reported
interaction becomes an edge; edges are indistinguishable; and the picture asserts
uniform confidence in claims whose evidential standing differs by orders of magnitude.
Filtering by experimental system or by throughput helps, but both are per-record
properties, and the question ``how well supported is \emph{this interaction}'' is a
property of the set of records supporting it.

ProLiVis 1.0 \cite{prolivis1} proposed treating the literature as the primary object of
visualization rather than as metadata: publications and experimental methods become
nodes in their own right, arranged in a three-level hierarchy around an organism. The
idea addressed the right problem. Its implementation had three limitations that this
paper removes.

\begin{enumerate}
  \item Publications were attached to a single, arbitrarily chosen experimental method.
        The grouping query selected a non-aggregated column, so a publication reporting
        three assays was filed under whichever the query planner returned first.
  \item Layout was delegated to a separate force-directed executable. Identical input
        produced different pictures on different runs, and the executable is no longer
        available, so the published figures cannot be regenerated.
  \item Evidence was displayed but never weighed. The tool could show that a
        publication reported an interaction; it could not say what that was worth.
\end{enumerate}

\subsection{Contributions}

\begin{itemize}
  \item A seven-term citation-trust model (\S\ref{sec:trust}) combining evidence
        internal to BioGRID with external bibliometric data. Its distinguishing term
        estimates the number of independent research groups behind an interaction by
        clustering the supporting publications over shared institutional identifiers.
  \item A deterministic center layout (\S\ref{sec:layout}) in which angular position
        encodes experimental method and radial position encodes hierarchy level, with
        closed-form placement and no iterative relaxation; and filters over the
        literature it draws --- contribution, proteins touched, method --- which
        recompute the method ring from the publications that survive.
  \item Structural extraction under an evidence threshold (\S\ref{sec:structure}),
        including edge removal in ascending order of trust, which asks what network
        structure survives if only well-supported interactions are believed.
  \item A recursive high-level reading of a network too large to draw
        (\S\ref{sec:structure}): modules as nodes and evidence as links, with any
        module openable into its own high-level graph. The grouping is chosen per
        level, because the structural decomposition that works on a sparsely studied
        organism returns a densely studied core unchanged.
  \item A zero-installation implementation (\S\ref{sec:implementation}) that ingests
        full BioGRID releases in the browser, keeps user data local, and emits session
        manifests sufficient to reproduce any figure.
\end{itemize}

\section{Related work}

\textbf{Network visualization.} Cytoscape \cite{cytoscape} is the standard platform for
biological network analysis; Gephi and Graphia serve general graph visualization. All
take interactions as given and provide no vocabulary for evidence quality beyond
whatever attributes the input file carries. The earlier PPI-specific tools surveyed by
\cite{prolivis1} --- ProViz, Osprey, Medusa, PIVOT, Robinviz --- share this property.

\textbf{Confidence scoring.} STRING \cite{string} assigns confidence scores by combining
evidence channels, but its experimental channel is a benchmarked probability derived
from co-complex association, not an account of how many independent groups reported a
given interaction. The IMEx consortium's MIscore \cite{miscore} scores interactions from
number of publications, interaction detection method and interaction type --- the
closest existing work to ours, and the natural comparison. MIscore's publication term is
a count; ours is a count of \emph{groups}, and we add terms for bibliometric impact and
recency. We regard MIscore as a baseline rather than a competitor and report against it
in \S\ref{sec:evaluation}.

\textbf{Literature-centric views.} Making publications first-class nodes is unusual.
The nearest neighbours are citation-network tools, which visualize papers without the
interactions they report, and provenance-aware curation interfaces, which record
evidence without drawing it.

\section{The citation-trust model}
\label{sec:trust}

Let $e$ be an undirected protein pair. Write $R(e)$ for the BioGRID records supporting
it, $P(e)$ for the distinct publications among them, and $S(e)$ for the distinct
experimental systems. Each term below maps to $[0,1]$, or to $\bot$ (unknown) when its
input is unavailable.

\subsection{Terms}

\paragraph{Replication.}
\begin{equation}
  t_{\mathrm{rep}}(e) = 1 - \exp\!\left(-\frac{|P(e)| - 1}{\kappa_p}\right),
  \qquad \kappa_p = 1.5 .
\end{equation}
Zero at $|P| = 1$: a single report is a claim. Saturating, so the second publication
contributes far more than the tenth.

\paragraph{Independence.} Let $\sim$ be the relation on $P(e)$ holding between two
publications that share at least one institutional identifier, and let $L(e)$ be the
set of equivalence classes of its transitive closure --- computed by union--find over
ROR identifiers obtained from OpenAlex \cite{openalex}, falling back to the first-author
label where affiliations are unresolved. Then
\begin{equation}
  t_{\mathrm{ind}}(e) = 1 - \exp\!\left(-\frac{|L(e)| - 1}{\kappa_l}\right),
  \qquad \kappa_l = 1.2 .
\end{equation}
This is the term a publication count cannot express. Consider two interactions each
supported by five publications: one by five separate laboratories, the other by one
laboratory five times. Their replication terms are identical by construction; their
independence terms are $t_{\mathrm{ind}} \approx 0.96$ and $t_{\mathrm{ind}} = 0$. The
term returns $\bot$ when no provenance is resolvable, rather than asserting that one
publication implies one group.

\paragraph{Method diversity.} Assays are partitioned into evidence classes --- binary,
co-complex, structural, proximity, enzymatic, colocalization, genetic --- grouped by
shared failure mode. Writing $C(e)$ for the classes spanned,
\begin{equation}
  \begin{split}
    t_{\mathrm{div}}(e) &= 1 - \exp\!\left(
      -\frac{(|C(e)| - 1) + \gamma\,(|S(e)| - |C(e)|)}{\kappa_d}\right), \\
    \gamma &= 0.3, \quad \kappa_d = 1 .
  \end{split}
\end{equation}
Two affinity-capture experiments can be misled by the same sticky bait; an assay from
another class fails differently. Repeating within a class therefore earns only partial
credit $\gamma$ relative to spanning classes.

\paragraph{Method directness.} $t_{\mathrm{dir}}(e) = \max_{s \in S(e)} d(s)$, where
$d$ is a per-assay prior on how directly the assay demonstrates physical contact,
ranging from $1.0$ for co-crystal structure to $0.2$ for co-localization and $0$ for
every genetic system. Assays absent from the bundled vocabulary receive $0.4$, so a
newly introduced high-throughput method cannot inflate scores before classification.

\paragraph{Throughput.} The low-throughput share of $R(e)$; $\bot$ when untagged.

\paragraph{Literature impact.} For publication $p$ with $c_p$ citations and age $a_p$
years, let $r_p = \log(1 + c_p / a_p)$. Then $t_{\mathrm{imp}}(e)$ is the empirical
percentile of $\max_{p \in P(e)} r_p$ within the dataset's corpus of citation rates. A
rate rather than a count, so recent work is not penalised for its age; a percentile
within the corpus, because absolute citation counts are not comparable across fields;
the maximum rather than the mean, so one landmark paper is not diluted by routine
papers that also mention the interaction. Unresolved publications yield $\bot$, never
$0$: scoring our own ignorance as absence of impact would systematically penalise older
and less-indexed literature.

\paragraph{Currency.} $t_{\mathrm{cur}}(e) = 2^{-\Delta/25}$ where $\Delta$ is years
since the most recent supporting publication.

\subsection{Combination}

Given weights $w_i$, the score is a weighted mean over the \emph{informed} terms, with
the weights of unknown terms redistributed:
\begin{equation}
  \mathrm{trust}(e) = \frac{\sum_{i : t_i \neq \bot} w_i\, t_i}{\sum_{i : t_i \neq \bot} w_i},
  \qquad
  \mathrm{coverage}(e) = \frac{\sum_{i : t_i \neq \bot} w_i}{\sum_i w_i}.
\end{equation}
Treating $\bot$ as a mid-range value would let missing data pull every score toward the
mean and make an unenriched dataset indistinguishable from a genuinely ambiguous one.
Coverage is reported alongside every score and included in every export: a score of
$0.6$ at $40\%$ coverage is a different claim from the same score at full coverage.

Three weight presets are shipped: \texttt{literature-aware} (default),
\texttt{evidence-only} (offline; bibliometric terms zeroed), and
\texttt{structural-strict} (directness-dominated). Weights are a documented default,
not a fitted optimum, and are editable.

\section{Center Layout 2.0}
\label{sec:layout}

\begin{figure*}[t]
  \centering
  \includegraphics[width=0.62\textwidth]{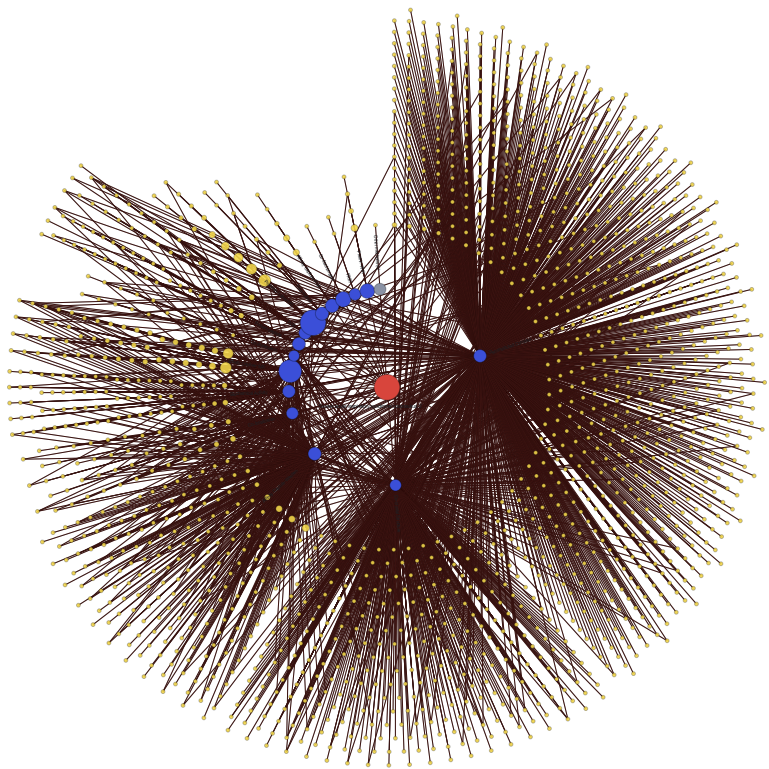}
  \caption{Center Layout 2.0 for SARS-CoV-2 (BioGRID 5.0.260): 1{,}688 publications
  across 19 experimental methods, with methods supported by fewer than five
  publications folded into a single node. Sector width encodes share of the literature;
  node area encodes interactions contributed. The two encodings disagree, and
  informatively so: the widest sectors (Reconstituted Complex, Co-crystal Structure)
  are where most \emph{papers} are, while the largest nodes (Affinity Capture-MS,
  Proximity Label-MS) are where most \emph{interactions} are.}
  \label{fig:center}
\end{figure*}

The layout places three levels of node on concentric rings: the organism at the origin,
experimental systems on the first ring, publications on an outer band
(Figure~\ref{fig:center}).

\paragraph{Sector allocation.} Each system $s$ receives an angular sector of width
proportional to $\max(\pi_s, \phi)$, where $\pi_s$ is its share of publications and
$\phi = 0.022$ is a floor ensuring that a method used by a single paper remains a
selectable target; widths are renormalized after flooring.

\paragraph{Publication placement.} A publication using a single method is assigned to
that method's sector. A publication using several is assigned to the sector nearest the
circular mean of its methods, computed by averaging unit vectors --- arithmetic mean of
angles is wrong on a circle --- and receives an edge to \emph{every} method it used.
This is the correction to the defect described in \S1: one publication is one node, and
its methods are all represented.

Within a sector, publications are packed into rows filling the band, with the capacity
of row $k$ at radius $\rho_k$ given by $\lfloor \theta_s \rho_k / \sigma \rfloor$ for
sector width $\theta_s$ and slot size $\sigma$. Row spacing shrinks so that the
outermost row remains inside a fixed band. Both properties are necessary at real scale:
letting rows march outward without bound causes a method with a thousand publications
to push its fan to several times the method ring's radius, collapsing every other method
into an unreadable sliver.

\subsection{Filtering the literature}
\label{sec:filter}

The band is bounded, so a literature larger than it holds is not drawn in full. In
release 5.0.260 the human literature is \HumanPublications{} publications, of which
\HumanPublicationsDrawn{} fit; the rest are omitted in ascending order of contribution.
No spacing draws forty thousand labelled nodes legibly, so the question is not whether
to omit but how the omission is chosen and whether the reader is told. Ours keeps the
largest contributors, and the interface reports what it drew against what exists.

The reader's own answer is to say which part of the literature they mean. Every
publication is one node here whatever it reported, and the quantity that separates a
screen from a structure paper is not on the page, so we filter on it directly: the
interactions a publication contributed, the distinct proteins it touched, and the
methods it used. The first two are not interchangeable --- ten interactions among three
proteins is a complex, ten among twenty is a screen --- and publications with identical
interaction counts routinely differ in protein count, so both are offered.

Filtering recomputes the method ring over the surviving publications rather than the
whole dataset. A sector's width is its share of the literature; a sector sized by papers
that are no longer drawn would describe a picture other than the one on screen. The
filter is recorded in the session manifest for the same reason: without it the manifest
would faithfully reproduce a different figure.

\begin{figure*}[t]
  \centering
  \includegraphics[width=0.60\textwidth]{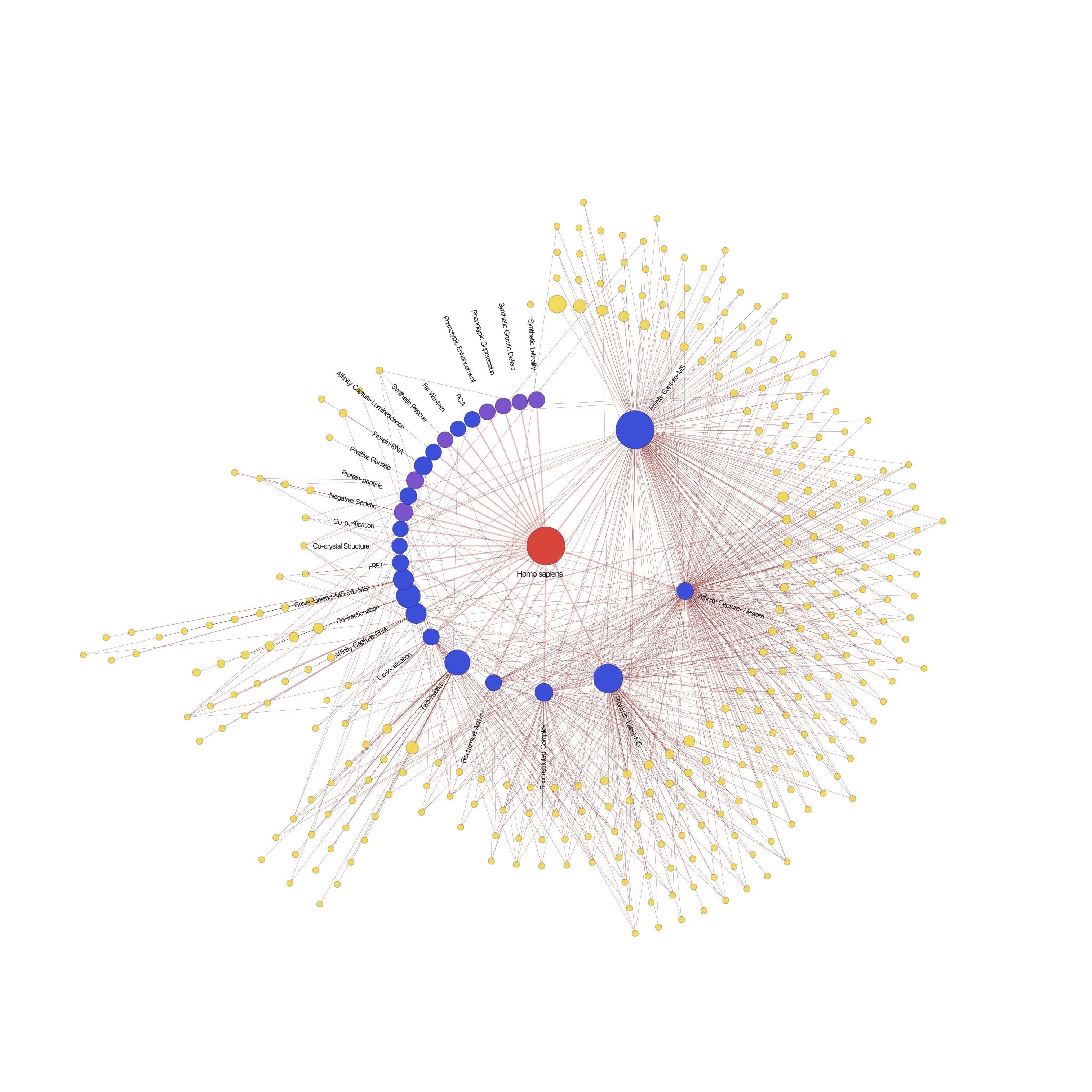}
  \caption{The human literature of BioGRID 5.0.260 filtered to the
  \ScreenPublications{} publications contributing at least \ScreenCut{} interactions
  each, drawn across \ScreenMethods{} methods. Unfiltered the same view holds
  \HumanPublications{} publications, of which \HumanPublicationsDrawn{} fit the band ---
  a picture in which no individual paper can be found. Publication labels are omitted
  here; at this count they collide.}
  \label{fig:screens}
\end{figure*}

Figure~\ref{fig:screens} is the human literature restricted to publications
contributing at least \ScreenCut{} interactions --- \ScreenPublications{} of the
\HumanPublications{}, across \ScreenMethods{} of the \HumanMethods{} methods. It is the
same layout, the same organism and the same options as the unreadable whole.

\paragraph{Determinism.} Inputs are sorted canonically, ties break on identifier, and no
random source or clock is consulted; string comparison is explicitly locale-independent,
since \texttt{localeCompare} would otherwise permit two machines to order method names
differently. Identical input therefore yields byte-identical coordinates. Complexity is
$O(n \log n)$, dominated by the sort.

\begin{figure*}[t]
  \centering
  \includegraphics[width=0.46\textwidth]{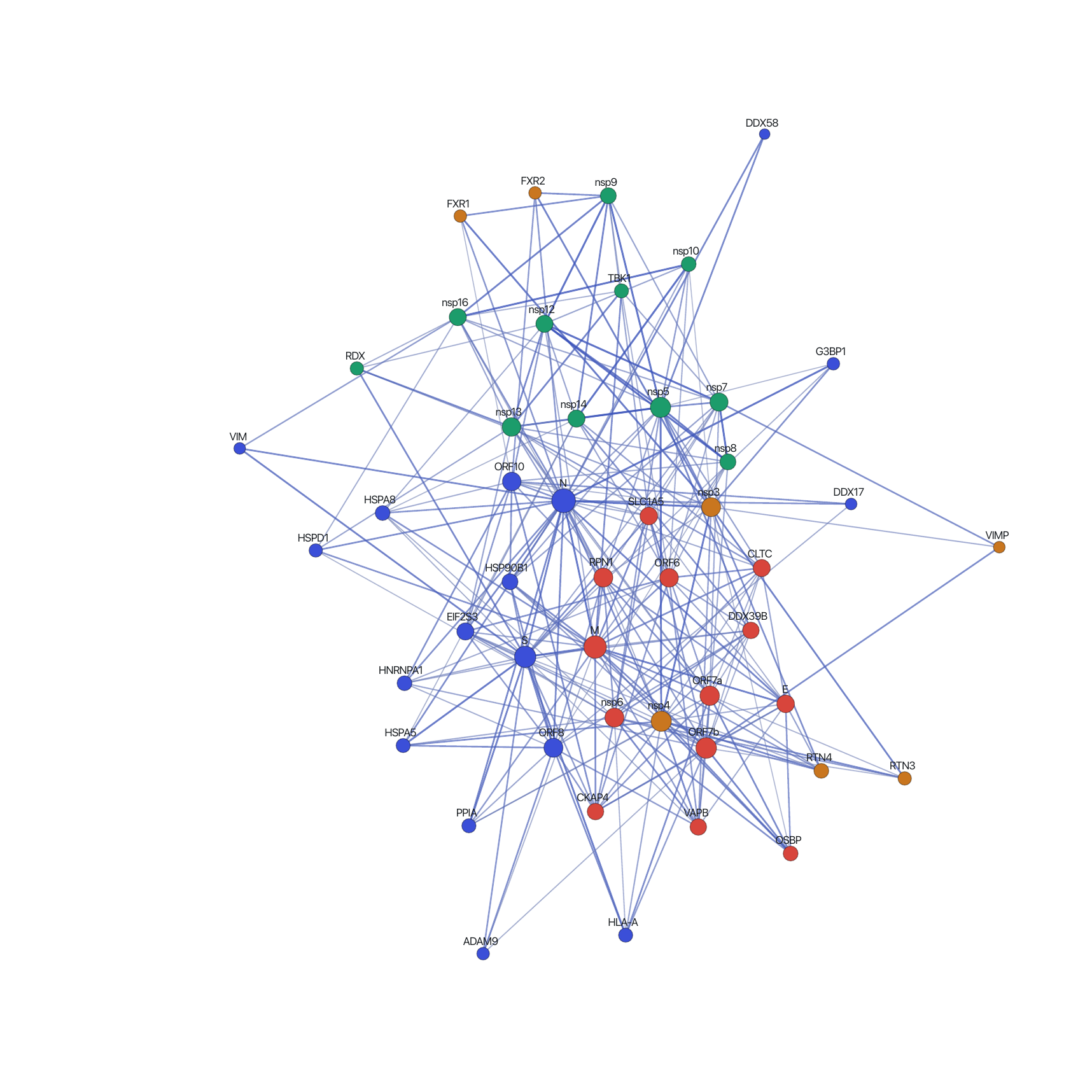}
  \hfill
  \includegraphics[width=0.46\textwidth]{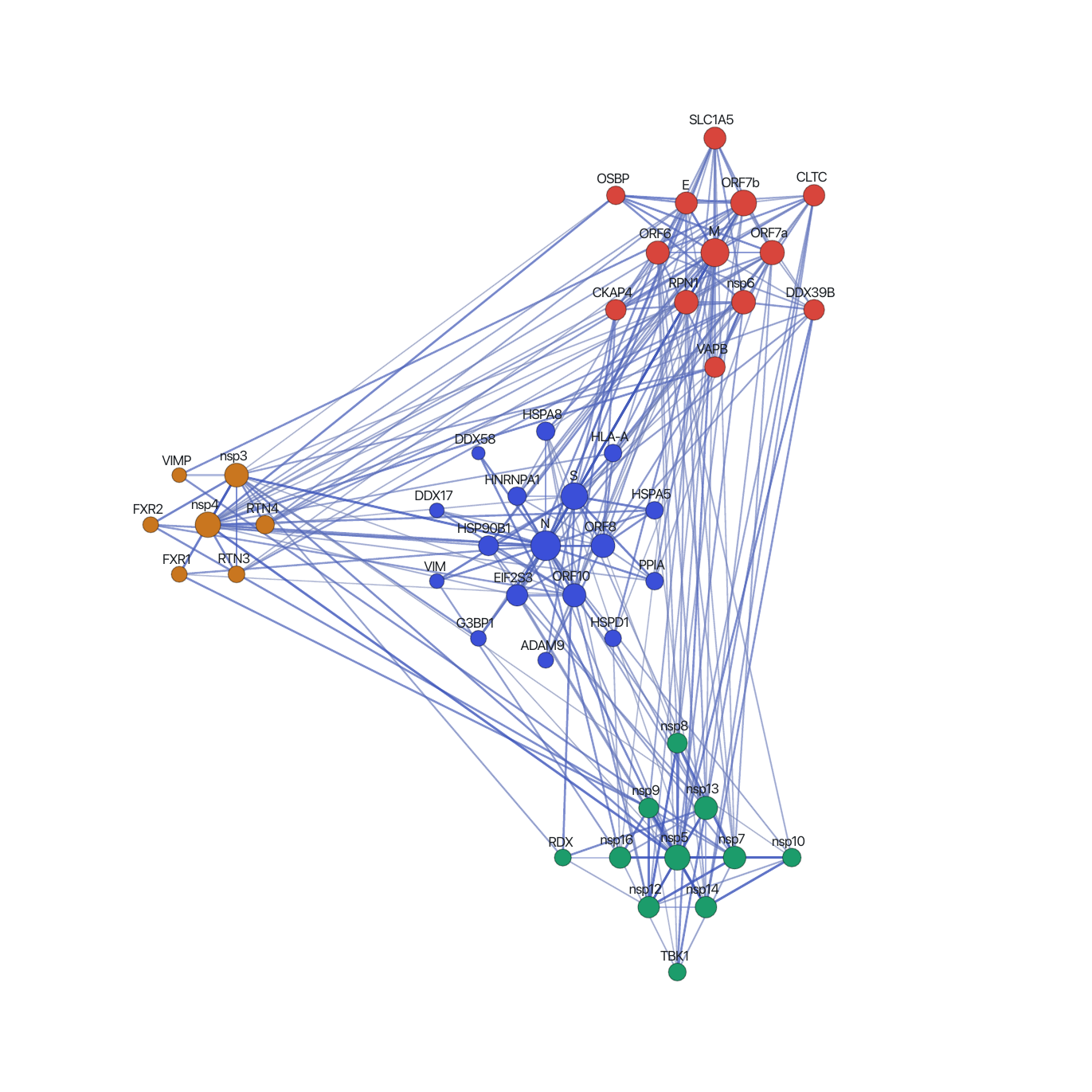}
  \caption{The same network under the two arrangements that answer different
  questions: SARS-CoV-2 in release 5.0.260 at trust $\geq 0.4$, keeping proteins with
  at least two partners and the best-supported \NetInteractions{} interactions ---
  \NetProteins{} proteins. Colour is the trust-weighted community. \emph{Left}, force:
  what clusters together. \emph{Right}, grouped: the \NetCommunities{} communities as
  packed discs. The communities are not told any biology, and correspond to it anyway
  --- the replication--transcription complex (nsp5, nsp7--nsp16) in green; the membrane
  and ER proteins M, E, ORF6, ORF7a/b and nsp6 with host VAPB, OSBP, CLTC and CKAP4 in
  red; nsp3 and nsp4 with the reticulons RTN3 and RTN4 in orange; N and S with host
  chaperones and RNA-binding proteins in blue.}
  \label{fig:network}
\end{figure*}

\section{Structural extraction under an evidence threshold}
\label{sec:structure}

Trust scores are edge weights, so every standard decomposition becomes available as a
function of how much evidence one demands: maximal cliques by Bron--Kerbosch with
pivoting over a degeneracy ordering, biconnected components and articulation points by
Hopcroft--Tarjan, $k$-cores, and contraction to a high-level graph whose inter-module
edges carry the summed trust of the original edges spanning them.

\subsection{Reading a network one level up}
\label{sec:highlevel}

A network of $10^5$ interactions cannot be read as a node--link diagram at any layout
quality. Contraction answers this only if it can be applied repeatedly: the modules of
a large network are themselves large. We therefore define a drill-down in which a
high-level node expands into a high-level graph of its own, and iterate until a level
is small enough to draw as proteins.

Which grouping to contract by cannot be fixed in advance. Biconnected components are
the natural first choice, being structural rather than optimised --- a module is a set
of proteins that remains connected when any one of them is removed --- and on a
sparsely studied organism they carry most of the answer, since the reported network is
largely bridges (\S\ref{sec:evaluation}). But a well-studied core \emph{is}
biconnected: the decomposition returns it unchanged, and a drill-down built on it
descends forever into the same picture. We therefore contract by biconnected components
when they decompose the graph and by modularity otherwise, using Louvain
\cite{louvain} weighted by trust, so that a community is held together by evidence
rather than by edge count. Modularity always divides, and divides again, which is the
property the recursion needs.

``Decomposes'' has to be stated carefully, and the human interactome shows why. Its
biconnected decomposition is not degenerate in the obvious sense --- it returns
thousands of components, and its largest holds 76\% of the proteins rather than all of
them --- but the remainder are two-protein bridges, so the drawing is one huge node
ringed by specks. We therefore require a \emph{second} module worth drawing: more than
one component, none holding more than 90\% of the proteins, and a second-largest
holding at least 1\% of the proteins and at least three of them.

Which grouping this selects is itself a measurement, reported in
\S\ref{sec:evaluation}: the structural decomposition is chosen for barely-studied
networks, whose reported interactions are little more than a forest of bridges, and
modularity for everything that has been studied enough to have a core.

\begin{figure*}[t]
  \centering
  \includegraphics[width=0.88\textwidth]{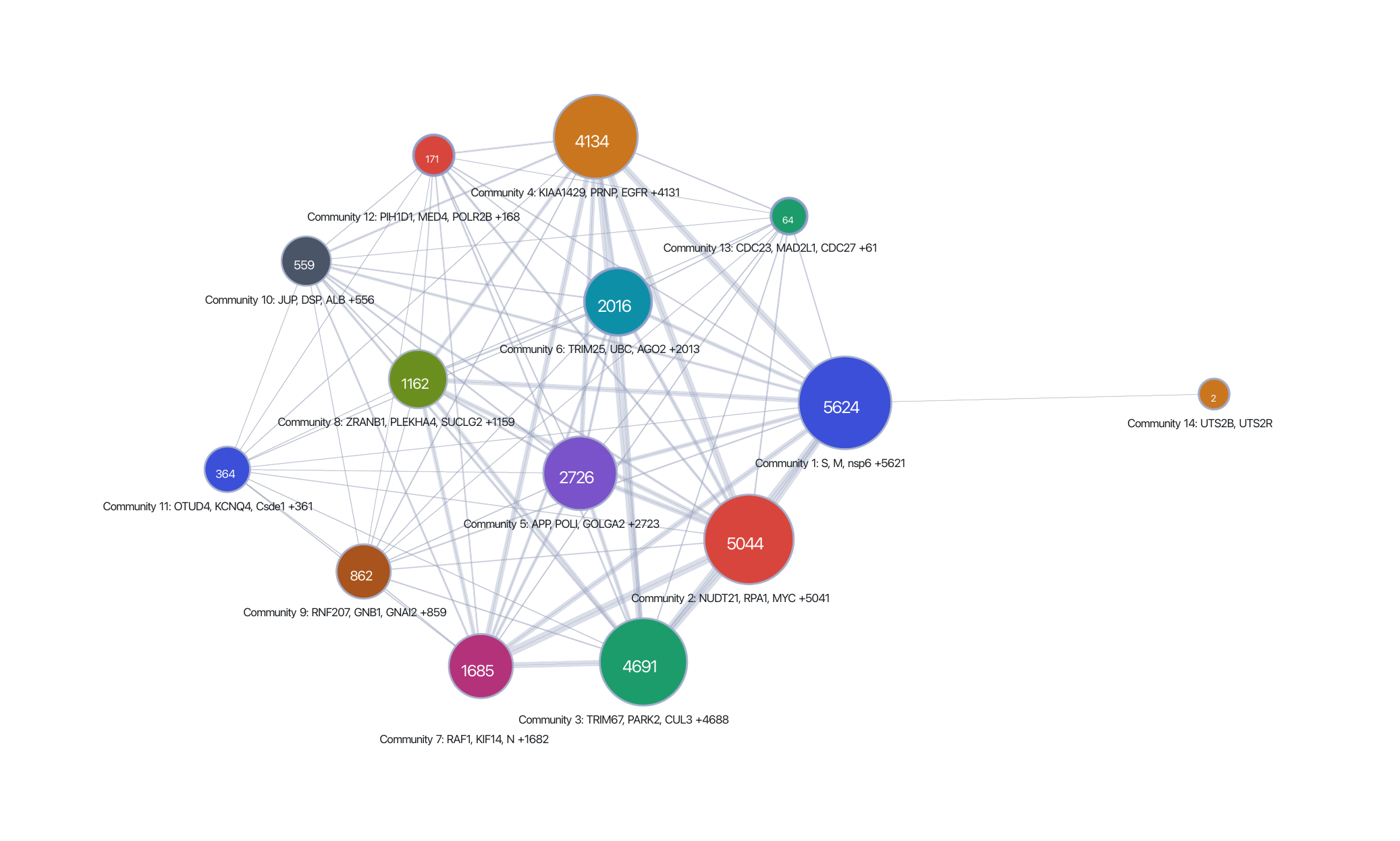}
  \caption{The complete human interactome of BioGRID 5.0.260 --- \HumanProteins{}
  proteins and \HumanInteractions{} interactions, no trust threshold --- read one level
  up, giving \HumanModules{} communities. Node area is proteins in the module, the
  number inside is that count, the ring is the mean trust of the module's internal
  interactions, and link width is the number of interactions spanning two modules.
  Community 14 sits apart because almost nothing connects it, which is a property of the
  network and not of the layout; how far apart is not, since a module thrown to the far
  distance by one weak spring is pulled back to the edge of the drawing.}
  \label{fig:highlevel}
\end{figure*}

\begin{figure*}[t]
  \centering
  \includegraphics[width=0.74\textwidth]{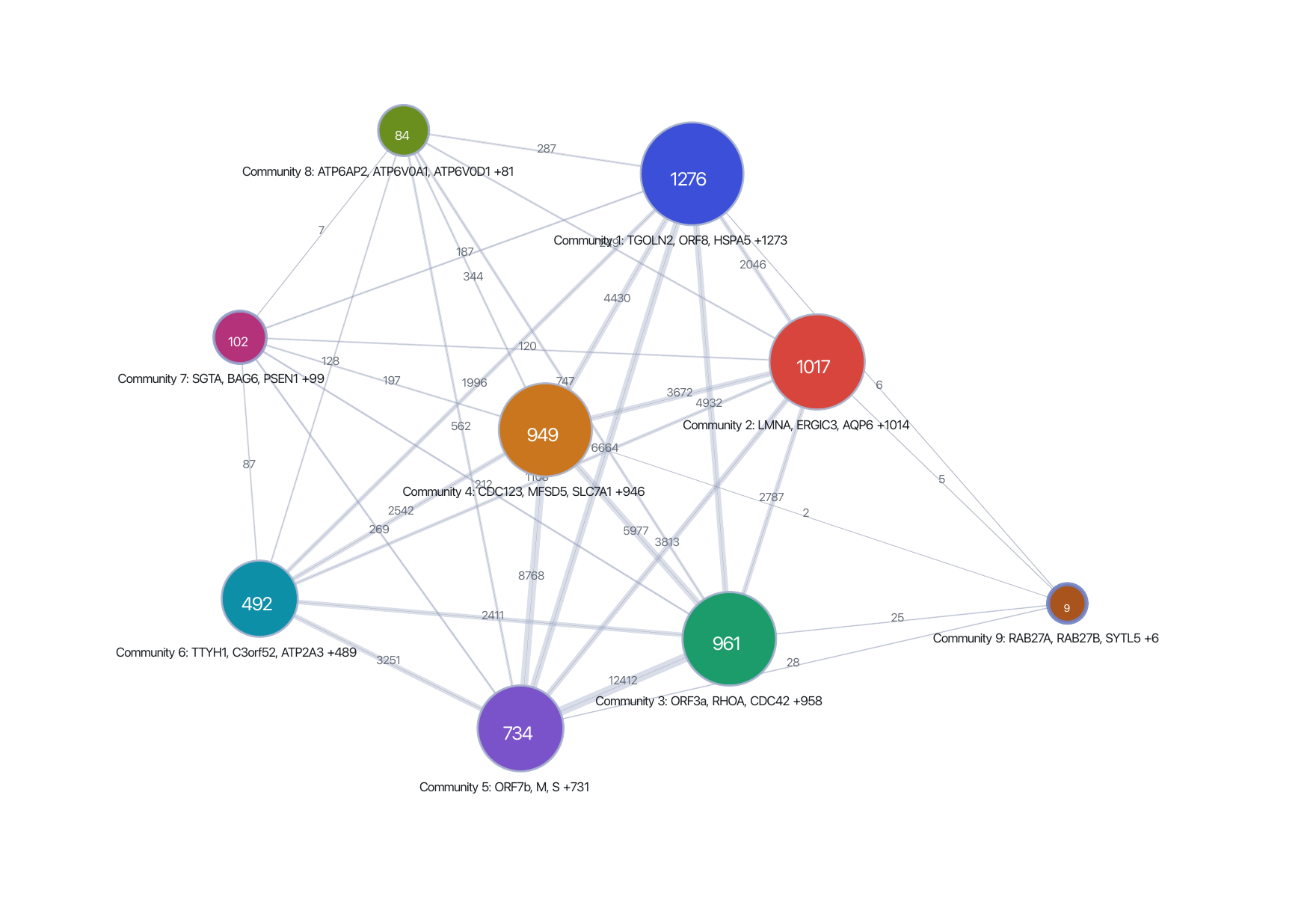}
  \caption{Community 1 of Figure~\ref{fig:highlevel} opened: 5{,}624 proteins and
  141{,}011 interactions, contracted again into nine communities. Link labels are
  interaction counts. The membership is coherent without having been told anything
  about biology --- a vesicular-transport and ER community (TGOLN2, HSPA5), a
  v-ATPase community (ATP6AP2, ATP6V0A1, ATP6V0D1), a small RAB27A/RAB27B/SYTL5
  secretory community, and two communities carrying the SARS-CoV-2 proteins that this
  organism's records interact with.}
  \label{fig:highlevel-drill}
\end{figure*}

Two details matter for correctness of the picture rather than of the algorithm.
Biconnected components share their articulation points, so the groups are made disjoint
before drawing --- each protein is claimed by the largest group containing it --- since
a protein drawn in two modules is counted twice, sized twice, and ambiguous to click.
And a long tail of two-protein modules is folded into a single node rather than
discarded, so no protein leaves the picture as one descends. Our Louvain implementation
visits nodes in index order rather than at random, making the partition, and therefore
the figure, reproducible.

We add one non-standard variant. Girvan--Newman removal deletes the edge of highest
betweenness, identifying joins in the network \emph{as reported}. Removal in ascending
order of trust instead deletes the least well-supported edge first, and answers a
different question: what structure survives if only the evidence is believed. Both
record component count and modularity at each step.

\section{Implementation}

\begin{figure}[t]
  \centering
  \includegraphics[width=\columnwidth]{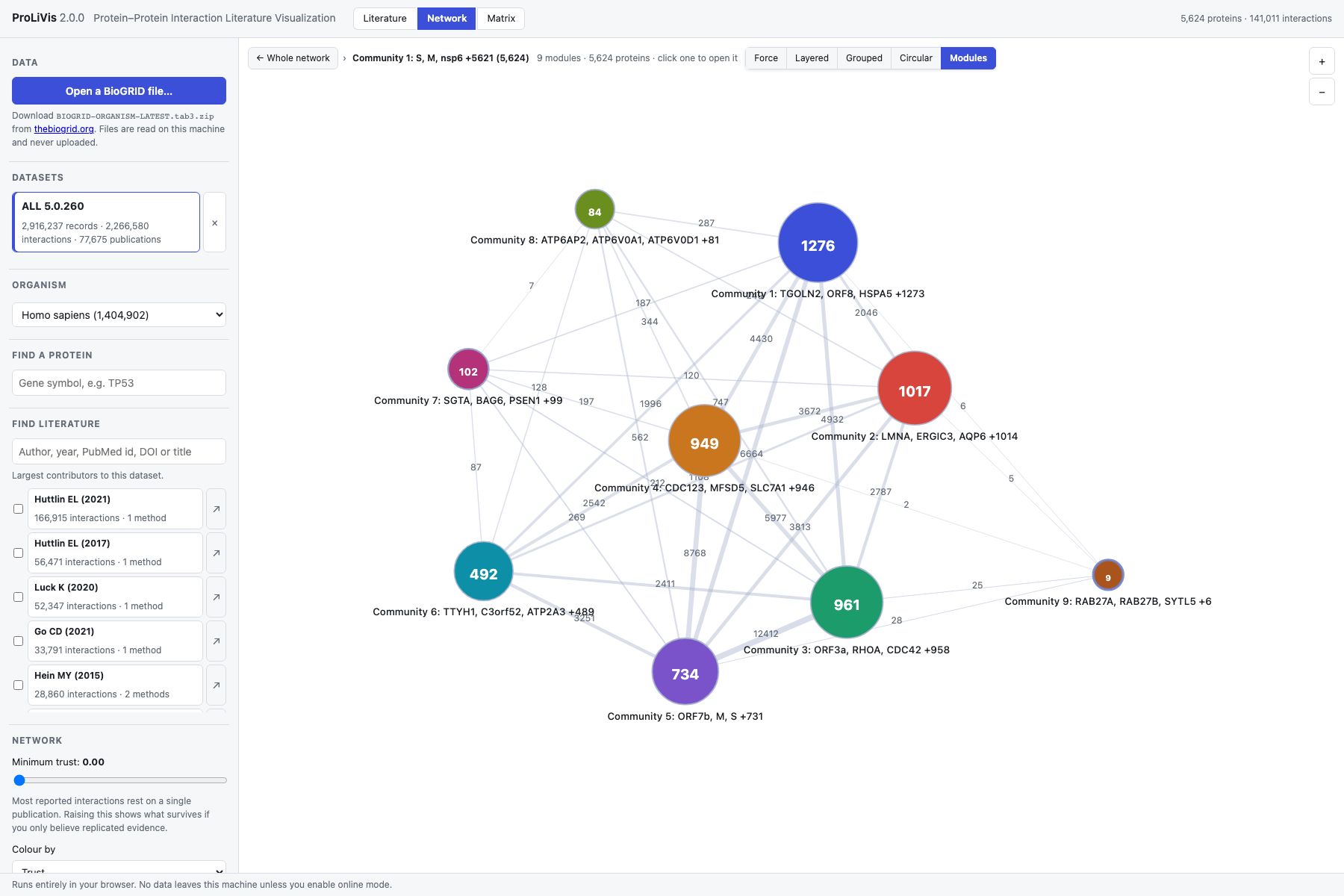}
  \caption{The interface, having opened one community of the human interactome. The
  breadcrumb above the canvas records the descent and returns to any level of it; the
  numbers on the links are interaction counts, and the panel on the left is the same
  trust threshold, density and literature search that apply to every view. The whole
  application is static files: this is a browser tab, with the release in an embedded
  analytical database beneath it.}
  \label{fig:interface}
\end{figure}
\label{sec:implementation}

ProLiVis 2.0 is a static web application: TypeScript, no server, no installation. Data
never leaves the user's machine unless online mode or bibliometric enrichment is
enabled, and both are optional.

Bulk BioGRID releases are ingested into DuckDB compiled to WebAssembly, persisted in the
browser's Origin Private File System. All parsing, coercion and aggregation happen in
the database; the application never iterates over interaction records in JavaScript.
Archive members are inflated in line-aligned chunks with the reader throttled, so peak
memory is bounded by one chunk regardless of archive size.

\paragraph{Drawing the network.} Repulsion in the force layout is exact below three
hundred proteins and Barnes--Hut above, which is what makes an organism-scale network
drawable at all: the exact computation takes 1.9\,s at a thousand proteins and
Barnes--Hut 0.3\,s, 0.6\,s at eighteen hundred, 2.6\,s at six thousand. Each connected
component is laid out separately and the components are then packed around the largest,
evenly around each ring. A component with no edge to the rest feels only repulsion from
it and otherwise drifts outward until gravity balances --- far outside everything else,
so that fitting the view to include it shrinks the part anyone came to see. Distance
between components therefore means nothing, which is honest: there is no interaction
between them for it to mean anything about.

The grouped arrangement uses the same trust-weighted communities as
\S\ref{sec:highlevel}, placed by the same packer. Grouping by connected component --- the
obvious choice, and ours at first --- does not work here: a PPI network is one component,
so every protein lands in one group and the arrangement degenerates to a single disc.
That is also what a force layout falls back to when a size cap defeats it, which is how
the two arrangements came to produce the same picture before Barnes--Hut removed the cap.
Figure~\ref{fig:network} is what they produce now.

Columns are resolved by alias onto a canonical set, because BioGRID's tabular formats
disagree: the bulk \texttt{tab3} files and the \texttt{tab2} output of the REST service
name six columns differently, and the REST service does not offer \texttt{tab3} at all.

Three properties of the data warrant recording, since each is a trap:
\begin{itemize}
  \item \texttt{Publication Source} is not always a PubMed identifier. Approximately a
        fifth of records in release 5.0.260 are DOI-referenced.
  \item \texttt{Throughput} is a set, not a scalar; a record may be tagged both high and
        low throughput.
  \item Gene symbols are not unique across organisms. In the coronavirus release alone,
        18 symbols map to more than one gene: \texttt{E}, \texttt{M}, \texttt{N} and
        \texttt{S} exist in three coronavirus species. Nodes are keyed by BioGRID gene
        identifier; symbol-keying would fuse the SARS-CoV-2 nucleocapsid with the
        SARS-CoV one.
\end{itemize}

Every figure carries a session manifest recording release, query, filters, trust
configuration and layout options. With a deterministic layout, this is sufficient to
regenerate the figure exactly.

\section{Evaluation}
\label{sec:evaluation}

\paragraph{Scalability.} On a consumer laptop, the complete BioGRID 5.0.260 release
(2{,}916{,}237 records, 1.55\,GB uncompressed) ingests in 78\,s, yielding 2{,}266{,}580
interactions across 98 organisms. The organism list then returns in 0.7\,s and the
experimental-system breakdown in 0.4\,s; the centre layout for \emph{Homo sapiens}
(16{,}178 nodes) takes 0.8\,s. Scoring all 1{,}068{,}827 human interactions takes
12\,s, after which re-scoring under a changed weight vector is pure arithmetic over
gathered evidence --- 1.6\,ms for 898 interactions --- which is what permits
interactive re-weighting. Maximal-clique extraction over the human graph thresholded at
$0.3$ (14{,}567 proteins, 122{,}317 interactions) takes 0.4\,s.

\paragraph{Reading a large network one level up.} Table~\ref{tab:drilldown} records the
descent through the complete human interactome of release 5.0.260 --- \HumanProteins{}
proteins and \HumanInteractions{} interactions, with no trust threshold applied ---
opening the largest module at each step, which is the worst case. \HumanSteps{} steps
take a reader from a million interactions to a module of \HumanLeaf{} proteins that is
drawn protein by protein, and the whole descent costs \HumanDescentSeconds{}\,s of
computation. Figures~\ref{fig:highlevel}
and~\ref{fig:highlevel-drill} are its first two levels.

\begin{table*}[t]
  \caption{Contracting the human interactome, opening the largest module each time.
  Generated by the figure pipeline rather than transcribed.}
  \label{tab:drilldown}
  \centering
  \small
\begin{tabular}{rrrlrrr}
  \toprule
  Level & Proteins & Interactions & Grouping & Modules & Largest & Time (s) \\
  \midrule
  0 & 29{,}104 & 1{,}047{,}820 & communities & 14 & 5{,}624 & 1.7 \\
  1 & 5{,}624 & 141{,}011 & communities & 9 & 1{,}276 & 0.2 \\
  2 & 1{,}276 & 15{,}102 & communities & 10 & 267 & 0.0 \\
  3 & 267 & 1{,}673 & communities & 8 & 53 & 0.0 \\
  \bottomrule
\end{tabular}

\medskip
{\footnotesize After the last step, 53 proteins remain, which is drawn protein by protein.}

\end{table*}

\paragraph{Which grouping a network needs.} Of the \OrganismsSurveyed{} organisms in
release 5.0.260 whose physical network has at least 100 proteins,
\OrganismsStructural{} decompose into biconnected components under the criterion of
\S\ref{sec:highlevel}; the rest are contracted by modularity. The split is not
arbitrary. The structural ones are the barely-studied networks --- \emph{Bos taurus}
(605 proteins, 586 interactions), \emph{Danio rerio} (529, 545), \emph{Gallus gallus}
(451, 472), Human Herpesvirus~1 (324, 395) --- where the reported network has scarcely
more interactions than proteins and is therefore close to a forest of bridges. The
well-studied ones --- human, yeast, mouse, fly, \emph{E.~coli}, \emph{Arabidopsis} ---
have a core that is biconnected by construction, and only modularity divides them. A
tool that offered biconnected components alone would work on the organisms nobody
needs help reading.

\paragraph{Filtering a literature too large to draw.} The human literature of release
5.0.260 comprises \HumanPublications{} publications across \HumanMethods{} methods, the
largest contributing \HumanBiggestPublication{} interactions on its own;
\HumanPublicationsDrawn{} of them fit the publication band. Restricting to publications
contributing at least \ScreenCut{} interactions leaves \ScreenPublications{} across
\ScreenMethods{} methods (Figure~\ref{fig:screens}), a picture in which individual
screens are identifiable. The filtered query costs 2.2\,s over the full release,
recomputing the method ring included.

\paragraph{Evidence structure.} Restricted to SARS-CoV-2 in release 5.0.260, the
physical network has \CovProteins{} proteins and \CovInteractions{} interactions in
\CovComponents{} connected component, of which \CovBridges{} interactions are bridges
and \CovArticulation{} proteins are articulation points. Raising the trust threshold to
\CovThreshold{} leaves \CovThresholdInteractions{} interactions among
\CovThresholdProteins{} proteins --- roughly a third. The reported network is a
well-connected core wrapped in a large periphery of claims that no second publication
has repeated.

\paragraph{Structural recovery.} Maximal-clique extraction over the same network finds
\CovCliques{} cliques of four or more proteins, the largest having \CovLargestClique{}
members: \CovLargestCliqueMembers{}. Without being told any biology, the largest cliques
are the SARS-CoV-2 replication--transcription complex --- nsp5, nsp9, nsp10, nsp12,
nsp13, nsp14, nsp15, nsp16 --- together with the host proteins that co-purify with it.
The caveat belongs with the result: these interactions come predominantly from
affinity-capture screens, in which a clique is evidence of co-purification and not of
simultaneous physical contact. The extraction reports what the literature asserts, which
is the same discipline the trust model observes.

\paragraph{Preset behaviour.} Over all \CovInteractions{} interactions,
\texttt{structural-strict} correlates with the assay-directness term at
$r = \CorrStrict$, against \texttt{evidence-only}'s $r = \CorrEvidence$ and
\texttt{literature-aware}'s $r = \CorrLiterature$. The presets therefore order
interactions differently rather than cosmetically, and in the direction their names
claim; none of them is a proxy for directness alone.

\paragraph{Discrimination.} \emph{To be completed.} The model should be evaluated as a
ranker against an external reference set it never saw --- CORUM co-complex pairs
\cite{corum} and hu.MAP \cite{humap} are the natural choices --- reporting AUROC and
average precision per preset and per organism, alongside MIscore \cite{miscore} as a
baseline, and a leave-one-term-out ablation. The harness for this is implemented and
tested; the numbers require running it against a reference set and are not reported
here. We state this rather than substituting an internally derived reference set, which
would be circular.

\section{Limitations}

The model scores evidence, not biology: a well-replicated artefact scores highly, and
should, because the problem lies in the literature and the score reports it faithfully.
It ranks interactions BioGRID already lists and never proposes new ones. It cannot see
co-citation bias; where a field converges on a belief, the bibliometric terms will
agree with the field. Institutional clustering is a proxy for laboratory identity, not
a measurement of it: two unrelated groups at one large university are merged, and a
group that has moved institutions is split. The default weights are reasoned, not
fitted, and \S\ref{sec:evaluation} is incomplete until they are tested against an
external standard.

\section{Availability}

Source, documentation and a hosted build are at
\url{https://github.com/melihsozdinler/CenterLayout} under the MIT licence. The
archived 1.0 implementation remains in the same repository under its original GPL-3
terms. All figures in this paper are generated by the tool from public BioGRID releases
by a single command.

\end{document}